\documentclass[twoside,leqno,twocolumn]{article}  
\usepackage{WZBP} 
\usepackage{cite}
\usepackage[pdftex]{color}
\usepackage[cmex10]{amsmath}
\usepackage{array}
\usepackage{mdwmath}
\usepackage{mdwtab}
\usepackage{enumerate}
\usepackage{eqparbox}
\usepackage{graphicx}
\usepackage[tight,footnotesize]{subfigure}
\usepackage{fixltx2e}

\usepackage{url}
\usepackage[noend]{algpseudocode}

\begin{document}
\title{\Large{Active Community Detection in Massive Graphs}
\thanks{This work is partially supported by the
XDATA program of the Defense Advanced Research Projects Agency (DARPA) administered through Air Force Research Laboratory contract FA8750-12-2-0303.}
}
\author{Heng Wang\thanks{Department of Applied Mathematics and Statistics, Johns Hopkins University, Email: hwang82@jhu.edu, cep@jhu.edu} \\
\and Da Zheng\thanks{Department of Computer Science, Johns Hopkins University, Email: dzheng5@jhu.edu, randal@cs.jhu.edu}\\
\and Randal Burns\footnotemark[3]\\
\and Carey Priebe\footnotemark[2]
}
\date{}

\maketitle
 

\begin{abstract} \small\baselineskip=10pt 
A canonical problem in graph mining is the detection of dense communities. This problem is exacerbated for a graph with a large order and size 
-- the number of vertices and edges -- as many community detection algorithms scale poorly. In this work we propose a novel framework for detecting active communities that consist of the most active vertices in massive graphs. The framework is applicable to graphs having billions of vertices and hundreds of billions of edges. Our framework utilizes a parallelizable trimming algorithm based on a locality statistic to filter out inactive vertices, and then clusters the remaining active vertices via spectral decomposition on their similarity matrix. We demonstrate the validity of our method with synthetic Stochastic Block Model graphs, using Adjusted Rand Index as the performance metric. We further demonstrate its practicality and efficiency on a most recent real-world Hyperlink Web graph consisting of over 3.5 billion vertices and 128 billion edges.
\smallskip


\end{abstract}


\section{Introduction}
There has been increased interest in community detection because communities in a large graph often imply noteworthy group structures in a graph-represented real system. For example, as summarized in \cite{fortunato2010community},  in a World Wide Web graph, communities are more likely to be groups of web pages associated with similar topics; in a Protein-Protein Interactions graph, communities are formed by proteins having the same functionality within a cell; in a scientific citation graph, communities are identified as research collaborators or potential collaborators. These valuable findings could further lead to concrete applications in business insights, security enhancement, recommendation systems and so on.  

To locate these communities, varied detection algorithms have been proposed.
Let $n$ denote the number of vertices of a graph and $m$ denote the number of edges. A traditional graph partitioning approach, the Kernighan-Lin algorithm \cite{kernighan1970efficient}, is still widely used today and has complexity $O(n^2\log n)$ and $O(n^2)$ on sparse graphs. A hierarchically agglomerative clustering approach, embedding all vertices in space to make use of a similarity measure, results in complexity $O(n^2)$ for single linkage and $O(n^2\log n)$ for a complete and average linkage scheme \cite{fortunato2010community}. A hierarchically divisive clustering algorithm proposed by Girvan and Newman \cite{girvan2002community} \cite{newman2004finding} iteratively partitions a graph by removing edges with low similarity and takes $O(nm^2)$.
In contrast, spectral clustering such as \cite{yang2008discovering} 
has much lower asymptotic computational complexity.
Their most expensive cost is to compute dominant Laplacian eigenvectors, which
has the complexity of $O(m)$ in each iteration but may require a large number of iterations\cite{Lanczos}.

Another prominent approach
is a group of modularity-based methods, developed from the stopping criterion of the Girvan and Newman algorithm in \cite{girvan2002community}. A greedy modularity optimization algorithm \cite{clauset2004finding} allows analysis of large graphs up to $n=10^6$ vertices with running time $O(n\log^2n)$ and is improved by \cite{wakita2007finding} to handle graphs up to $n=10^7$. In the past few years, the modularity-based technique \cite{blondel2008fast} known as Louvain clustering is in vogue because it can analyze graphs sizes up to $m=10^9$ in a reasonable time. The phase of attaining local modularity maxima in Louvain clustering requires multiple iterations and each iteration has complexity $O(m)$. The downside is that the number of iterations is unknown and convergence speed is influenced by the order of sequential sweeps over all vertices.

Almost all popular partitioning or clustering procedures are computed from the full topology of a graph and thus have high computation complexity.
It is challenging to run these algorithms on a billion-scale graph.
For example, the most recent Hyperlink Graph has $3.5$ billion and $128$ billion edges
\cite{hyperlinkgraph_source}, the largest graph available to the public.
Even growing at $O(m)$ in each iteration, Louvain clustering and spectral clustering potentially require many iterations to converge, which is computationally challenging to work at the billion scale,
let alone algorithms with the complexity of $O(n^2\log n)$ or $O(nm^2)$. 
Thus, it is important to consider
the situation where a graph is too large to be processed on its full topology.

Moreover, sometimes it is only dense and comparatively active groups of vertices that we are concerned with in graph analysis. Dense clusters consisting of only inactive vertices in a giant network,
e.g. small cliques incorporating only insignificant websites in the Hyperlink Graph,
are unimportant for observers. In this scenario, 
investigations solely on active vertices are sufficient to detect potential communities consisting of the most active vertices. In this paper, we propose to use a locality statistic \cite{wang2014} to measure the activity level of a vertex. The communities that consist of the most active vertices are referred to as ``active communities". For example, some link farms in web graphs are ``active communities".


The contribution of this work mainly has two facets. Firstly, we propose an alternative community detection framework because it is unattainable to cluster on an entire massive graph due to the large graph order or size and it is only active vertices that are important in many networks. The framework identifies the most active vertices, i.e. the ones of the largest locality statistic values, builds a smaller graph over active vertices and then assigns the most active vertices into communities through typical clustering methods. Secondly, to unearth the most active vertices in a network, we provide a highly parallelizable trimming algorithm to screen out inactive vertices. The number of discovered active vertices is much smaller than graph order $n$. We apply our methodology on the famous Hyperlink Graph \cite{hyperlinkgraph_source} to identify active communities. To the best our knowledge, this is the first community detection algorithm applied to a real graph dataset at this scale.

An outline of the paper is given as follows. Notations and Locality statistic will be elaborated in \S~\ref{sec:locality-intro}. 
\S~\ref{sec:detection-framework} presents procedures in our active community detection algorithm framework. \S~\ref{sec:local.scan-implement} describes a parallelizable trimming algorithm that cost-effectively skips actual computation on the majority of vertices. In the \S~\ref{sec:exp-synthetic}, our detection algorithm is empirically validated on graphs with true and known community structures. For the real data experiment in \S~\ref{sec:page-graph data},  we apply the proposed algorithm on the massive Hyperlink Graph collected recently in  \cite{hyperlinkgraph_source}. \S~\ref{sec:discussion} concludes the paper and future research direction in this area.

\section{Locality Statistic}
In this paper, we consider only directed and unweighted graphs without self-loops. All procedures can be easily adapted to undirected or weighted graphs if necessary (\S~\ref{sec:discussion}). Generally, a graph is denoted by $G$ with vertex set $V=V(G)$ and edge set $E=E(G)$. The list of incident edges of a vertex $v$ is denoted by $E[v]$. For any $u,v \in V$, we use 
$d(u,v)$ for the shortest path distance of $(u,v)$ on the underlying undirected graph after removing orientations of all edges. For $v \in V$, we denote by $N_{k}[v]$ the set of vertices $u$ at distance at most $k$ from $v$, i.e., $N_{k}[v]= \{u \in V \colon d(u,v) \leq k \}$. For $V' \subset V$, $\Omega(V',G)$ is the subgraph of $G$ induced by $V'$. Thus, $\Omega(N_k[v],G)$ is the subgraph of $G$ induced by vertices at distance at most $k$ from $v$.
			
\label{sec:locality-intro} 
Locality statistic is commonly used in graph mining to detect a local region in the graph with significantly excessive intra-region connections \cite{wang2014}. The locality statistic of some vertex $v$ is the number of edges within the $k$-th order neighborhood of $v$. $k$ can be seen as the implicit and limited horizon that a vertex often reaches within the network.
Large locality statistic foreshadows the existence of a dense $k$-th order neighborhood centering at $v$. Hence, the locality statistic of a vertex becomes a measure of activity level of the vertex in the network. 

\begin{figure}
		\vspace{-2mm}
		\hbox{\hspace{0em}
	 	 \includegraphics[scale=0.7]{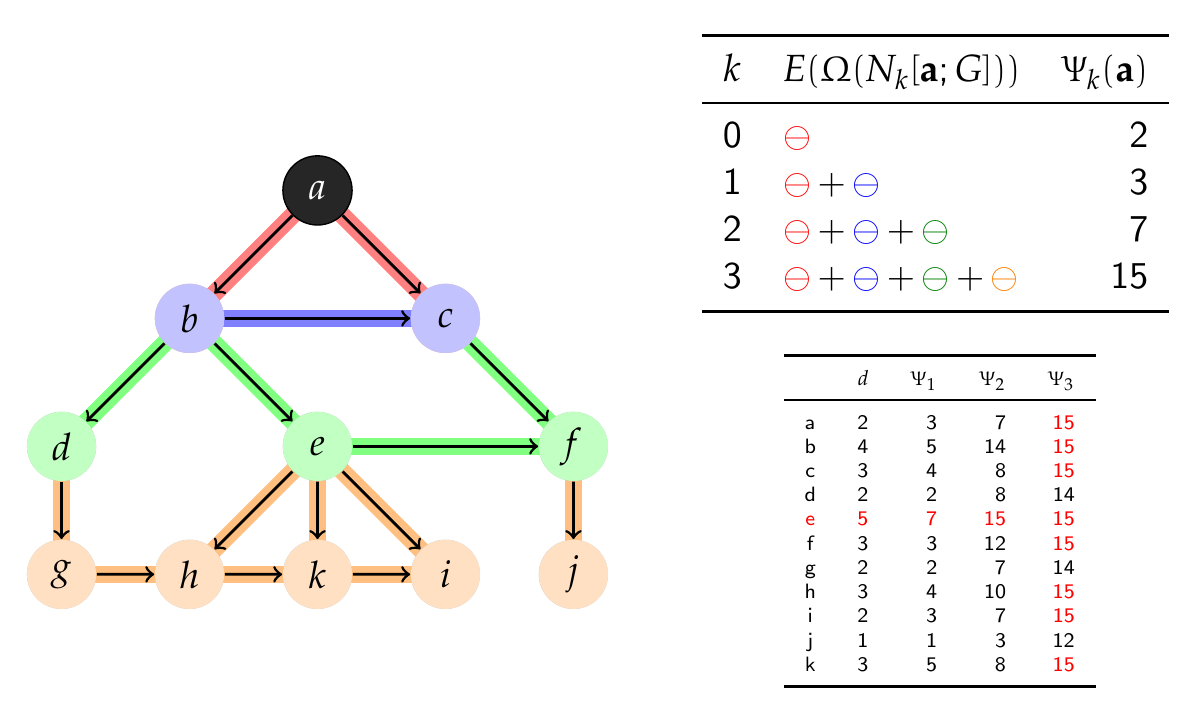}}
		\vspace{-3mm}
	 	 \caption{A toy example to illustrate calculations of $\Psi_{k}(a)$ with various $k=0,1,2,3$, on the directed $G$. For example, if $k=2$, $N_{2}[a]= \{u \in V \colon d(u,a) \leq 2 \}=\{a,b,c,d,e,f\}$ and thus $E(\Omega(N_{2}[a],G))$ contains edges colored in 					red, blue and green.}
	 	 \label{fig:example-of-locality}
\end{figure}

Formally, let $G$ be a graph. The locality statistic $\Psi_{k}(v)$ for all $k \geq 1$ and $v \in V$ on $G$ is defined as 
	\begin{equation} 
	\label{def:locality_stat} 
	\Psi_{k}(v)=|E(\Omega(N_{k}[v],G))|. 
	\end{equation}
Since $G$ is unweighted, $\Psi_{k}(v)$ counts the number of edges in the subgraph of $G$ induced by $N_{k}[v]$, a local territory of $v$ where all elements are at a distance at most $k$ from $v$ in $G$. More specifically, if $k=1$, the case thoroughly investigated in this paper, $\Psi_{1}(v)$ counts the number of edges either incident to $v$ or involved in triangles containing $v$. Large locality statistic implies a dense region whose members are all inclined to be ``friends" with each other, and such a region is not necessarily limited to a clique.
If we use a clique to locate a dense region, a subgraph $N_{k}[v]$ such as the one with all possible internal links except one is undervalued even though it is an extremely cohesive region. In a slight abuse of notation, we let $\Psi_{0}(v)$ be the sum of in-degree and out-degree of $v$. A simple toy example (\figurename~\ref{fig:example-of-locality}) illustrates calculations of $\Psi_{k}(a)$ with $k=0,1,2,3$, on the directed graph $G$.

\section{Algorithm}
The proposed detection framework is mainly composed of three steps and the first step is novel. The first part of this section elaborates the rational behind each step. As a key contribution of this work, the second subsection presents a trimming algorithm that efficiently identifies the most active vertices in a graph.
\subsection{Detection framework}
\label{sec:detection-framework}
Given a graph, our detection framework is summarized as follows:
\begin{enumerate} [(i)]
\item
Find the set of the top $Q$ most active vertices $\mathcal{C}$, i.e., the ones of the $Q$ largest locality statistic values $\Psi_k(v)$; we re-denote them as $\mathcal{C}=\{v_1,v_2,\dots,v_Q\}.$
\item
Construct a similarity matrix $\mathcal{S}$ for vertices in $\mathcal{C}$. $\mathcal{S}$ is a $Q\times Q$ matrix where $\mathcal{S}_{i,j}$ measures similarity between vertex $v_i$ and $v_j$. 
\item
Run a clustering algorithm on the similarity matrix $\mathcal{S}$ and report clusters as active communities. 
\end{enumerate}

Step (i) employs the locality statistic as a quantity to identify the $Q$ largest hubs in a network whose activity level is evaluated in the $k$-th order neighborhood.
Similar criteria of defining local activity level are proposed in \cite{bagrow2005local} \cite{eckmann2002curvature}: \cite{bagrow2005local} uses the L shell method to agglomeratively find a community for each vertex, which has extremely heavy computational burden in large graphs; \cite{eckmann2002curvature} distinguishes vertices based on high clustering coefficients, which may trigger false alarms on small cliques such as triangles in our problem. It is computationally expensive to compute locality statistic on all vertices. For example, if $k=1$, the computation on all vertices has 
a complexity $O(md_{max})$ where $d_{max}$ is the largest vertex degree in the graph. Thus, we deploy a trimming algorithm to 
obtain the top $Q$ largest locality statistic values, shown in \S~\ref{sec:local.scan-implement} for the case $k=1$.
 Also, note that in certain applications some of the most active vertices might be considered as outliers if they are not connected to the rest of top $Q$ vertices. These outliers are unnecessarily taken into account for community identification. In that case, a larger $Q$ should be used and the largest outlying vertices should be trimmed. We do not use this assumption in this paper but all methodologies and arguments can be easily adapted for this assumption.

After identifying the most active vertices, our framework uses the spectral clustering approach instead of popular modularity-based methods for several reasons. First of all, modularity optimization requires information of the whole graph so that information extracted solely from active vertices is less likely to be applicable to modularity-based methods. Furthermore, modularity maximum does not necessarily mean that a graph has a community structure and also high proximate modularities can fail to be similar partitions \cite{fortunato2010community}. These reasons lead us to construct a similarity matrix on $\{v_1, \dots, v_{Q}\}$.

In Step (ii), the problem of quantifying similarities between active vertices has received significant attention. Two main types of similarities have been studied in a large number of application domains: vertex feature based and network structure based. The former quantifies similarities resting on attribute values of each vertex such as \cite{Young07randomdot}. The latter focuses only on graph topologies: a pair of vertices achieve a high degree of similarity if they share many neighbors. Some classic measures can be used, such as \textit{Jaccard Index} 
$
	\label{jaccard_index}
	S_{ij}=|N_k[v_i]\cap N_k[v_j]|/|N_k[v_i]\cup N_k[v_j]|
$
where $N_k[v]$ is the set of vertices at distance at most $k$ from  $v$ in original graph $G$.
Assessment of similarities between active vertices $\{v_1,v_2, \dots,v_Q\}$ is not the main aim of this paper. Whether to select a feature based or network structure based approach or which classic measure to be used is application domain dependent.

Once the similarity matrix $\mathcal{S}$ is available, we can cluster the $Q$ vertices through a large number of standard clustering algorithms such as
Graph Spectral Clustering. In this paper, we prefer to cluster the $Q$ vertices
through spectral clustering \cite{ng2002spectral} on $S$ 
because the representation induced by eigenvectors enables the clustering distinctness of
initial data points to be more evident \cite{fortunato2010community}. With the spectral decomposition above,
the large gaps between consecutive eigenvalues suggest the number of clusters in a graph. 

\subsection{A trimming algorithm}
\label{sec:local.scan-implement}
\begin{figure}[tbh]
\begin{algorithmic}[1]
\Function {$local\_stat$}{$v$}
	\State $lstat \gets 0$
	\ForAll {$u \in N_{1}[v]$}
		\ForAll {$e \in E[u]$}
			\If {$S[e] \in N_{1}[v]$
				and $D[e] \in N_{1}[v]$}
				\State $lstat \gets lstat + 1$
			\EndIf
		\EndFor
	\EndFor
	\State \Return $lstat / 2$
\EndFunction
\end{algorithmic}

\begin{algorithmic}[1]
\Function {$est\_lstat1$}{$v$}
	\State \Return $\Psi_{0}(v)^2 + \Psi_{0}(v)$
\EndFunction
\end{algorithmic}

\begin{algorithmic}[1]
\Function {$est\_lstat2$}{$v$}
	\State $est \gets 0$\
	\ForAll {$u \in N_{1}[v]$}
		\State $est \gets est + min(\Psi_{0}(u), |N_{1}[v]| \times 2)$
	\EndFor
	\State \Return $est / 2$
\EndFunction
\end{algorithmic}
\caption{$local\_stat$($v$) computes $\Psi_{1}(v)$. $est\_lstat1$($v$)
	and $est\_lstat2$($v$) compute the upper bound of $\Psi_{1}(v)$.
	$est\_lstat2$($v$) computes a much tight upper bound but requires
	more expensive computation.
$S[e]$ denotes the source vertex of an edge $e$ and $D[e]$ denotes
the destination vertex of an edge $e$.}
\label{lstat}
\end{figure}

\begin{figure}[tbh]
\begin{algorithmic}[1]
\Function {$top\_lstat$}{$V$, $curr\_max$}
	\State sort $V$ s.t. degree(V) in DESC
	\State $V' \gets \{\}$
	\ForAll {$v \in V$}
		\State $est \gets est\_lstat1(v)$
		\If {$est \geq curr\_max$}
			\State $est \gets est\_lstat2(v)$
		\EndIf
		\If {$est \geq curr\_max$}
			\State $lstat \gets local\_stat(v)$
			\State $V' \gets V' \cup \{v\}$
			\State $curr\_max \gets max(lstat, curr\_max)$
		\EndIf
	\EndFor
	\State \Return $V'$
\EndFunction
\end{algorithmic}

\begin{algorithmic}[1]
\Function {$topQ\_lstat$}{$V, Q$}
	\State $curr\_max \gets 0$
	\State $knownV \gets \{\}$
	\While {$|knownV| < Q$}
		\State $V' \gets top\_lstat(V, 0)$
		\State $V \gets V \setminus V'$
		\State $knownV \gets knownV \cup V'$
	\EndWhile

	\State sort $knownV$ s.t. $local\_stat(V)$ in DESC
	\State $kth\_lstat \gets 0$
	\While {$kth\_lstat \neq local\_stat(knownV[Q])$}
		\State $kth\_lstat \gets local\_stat(knownV[Q])$
		\State $V' \gets top\_lstat(V, kth\_lstat)$
		\State $V \gets V \setminus V'$
		\State $knownV \gets knownV \cup V'$
		\State sort $knownV$ s.t. $local\_stat(V)$ in DESC
	\EndWhile
\EndFunction
\end{algorithmic}
\caption{$top\_lstat$ computes the largest locality statistic among a set
	of vertices $V$. $topQ\_lstat$ finds the vertices of $Q$ largest locality
statistic values among $V$.}
\label{top_scan}
\end{figure}

%

In the first step of our framework, we need to identify the vertices of the largest locality statistic values in a massive graph.
It is inefficient to compute locality statistic values of all vertices, while we only need to identify the largest ones.
As shown by $local\_stat$ in Figure \ref{lstat}, $\Psi_1(v)$ counts the number of edges,
of which adjacent vertices are both in $N_1[v]$, in the collection of incident edges of vertices in $N_1[v]$.
That is, $\Psi_1(v)=\dfrac{1}{2}\sum_{u\in N_1[v]}\sum_{e\in E[u]}\mathbf{1}_{\{S[e]\in N_1[v] \wedge D[e]\in N_1[v]\}}$,
where $\mathbf{1}_{\{\cdot\}}$ is an indicator function.
The complexity of computing $\Psi_1(v)$ of all vertices is $O(md_{max})$.

Therefore, we deploy a cost-effective trimming algorithm to safely skip the computation of $\Psi_{1}(v)$
on the vertices with small locality statistic, while still being able to identify the vertices with
the $Q$ largest locality statistic values. The trimming algorithm skips the wasteful computation
based on the upper bound of the locality statistic of a vertex. The tighter upper bound we achieve,
the more vertices on which we can skip computation. The procedures in the rest of the section describe
the trimming algorithm that works for the first-order neighborhood.  

We develop two upper bounds of $\Psi_{1}(v)$
in our trimming optimization, shown by $est\_lstat1(v)$ and $est\_lstat2(v)$ in Figure \ref{lstat}. 
$est\_lstat1(v)$ is a very loose but computationally efficient upper bound. Because $v$ has at most $\Psi_{0}(v)$ neighbors,
$\Psi_{1}(v)\leq \Psi_0(v)^2+\Psi_0(v)$ and the equality holds when all neighbors of $v$ are fully connected. $est\_lstat2(v)$ computes a much tighter upper
bound and is also more computationally expensive.
We denote by $contr_{v}(u)$ the amount of potential contribution of $u \in N_1[v]$ to $\Psi_{1}(v)$. 
$\Psi_{1}(v)$ is upper bounded by the sum of $contr_{v}(u)$ over all neighbors in $N_1[v]$, i.e. $\Psi_{1}(v)\leq \sum_{u\in N_1[v]}contr_{v}(u)$.
$contr_{v}(u)$ meets two inequalities: $contr_{v}(u) \leq \Psi_{0}(u)$ and $contr_{v}(u) \leq 2 \times |N_{1}[v]|$,
because the number of distinct directed triangles incorporating both $u$ and $v$ is upper bounded by $\Psi_0(u)$ and $2|N_{1}[v]|$.
Since $\sum_{u\in N_1[v]}contr_{v}(u)$ counts each potential edge twice, we divide the sum by two. Although $\dfrac{1}{2}\sum_{u\in N_1[v]}min(\Psi_{0}(u), |N_{1}[v]| \times 2)$ is not the tightest bound, it is sufficiently accurate to eliminate computation of locality statistic on most vertices.

Having upper bounds $est\_lstat1$ and $est\_lstat2$,
we now describe our procedure of finding $\arg\max_{v\in V} \Psi_{1}(v)$ over any set of vertices $V$, illustrated by $top\_lstat$ in
Figure \ref{top_scan}. The idea is to maintain the largest locality statistic discovered so far ($curr\_max$)
and skip expensive computation on the vertices whose upper bound of locality statistic
is smaller than $curr\_max$. Since $est\_lstat2$ 
requires more computation than $est\_lstat1$, we compute
$est\_lstat1$ first and only compute $est\_lstat2$ if $est\_lstat1$ is greater
than $curr\_max$.
To reach $\arg\max_{v\in V} \Psi_{1}(v)$ early, the procedure starts from the vertices with the largest degree with an assumption
that a larger-degree vertex is more likely to have a larger locality statistic.
To accelerate finding top $Q$ vertices, $top\_lstat$ returns not only $\arg\max_{v\in V} \Psi_{1}(v)$ but also all of
the vertices whose locality statistic has been computed during the process of finding $\arg\max_{v\in V} \Psi_{1}(v)$.


By utilizing $top\_lstat$, $topQ\_lstat$ (Figure \ref{top_scan}) finds vertices with the $Q$ largest locality statistic values.
$topQ\_lstat$ takes two stages to look for vertices with the largest locality statistic. In the first stage, we find at least $Q$ vertices of large locality statistic by repeatedly invoking $top\_lstat$
on the remaining vertices in the graph whose locality statistic is unknown. In the second
stage, we use $top\_lstat$ to continue searching for vertices with the largest
locality statistic among the remaining vertices in the graph whose locality
statistic is unknown. The procedure stops when $top\_lstat$ can no longer discover
a vertex whose locality statistic is larger than the current $Q$th largest locality
statistic.

The complexity of computing top $Q$ locality statistic values depends on both graph structures
and the parameter $Q$. Theoretically, a very loose upper bound of the complexity is
$O(md_{max})$, the complexity of computing locality statistic on all vertices.
However, its complexity in practice is much smaller when $Q\ll n$ because the trimming
algorithm skips computation on the majority of the vertices in a graph.
For example, if $Q=100,000$, our algorithm only needs to compute locality statistic
on $163,409$ vertices in the Hyperlink graph, which account for $0.0047\%$ of vertices in the graph.
The complexity of running
$est\_lstat1$ on all vertices is $O(n)$ and running $est\_lstat2$ on all vertices
is $O(m)$. Therefore, the complexity of the trimming algorithm throughout the entire
computation is between $O(n)$ and $O(m)$ where the constant factor here is $1$.

\subsubsection{Shared-memory parallel implementation}
In this section, we describe the parallel implementation of our algorithm
in shared memory. Although trimming skips unnecessary computation on many vertices to
speed up computation, a parallel implementation is still necessary for a graph
with billions of vertices, especially in the era of multi-core processors.
We implement our algorithm in FlashGraph \cite{FlashGraph}, a programming
framework for large-scale graph analysis. The implementation is written in C++.

We parallelize our implementation by parallelizing the function $top\_lstat$
since its computation on each vertex is independent. We split the vertices
in a graph into multiple partitions and create a thread for each partition
to process the vertices in the input set of $top\_lstat$ in parallel.
Once a thread completes all vertices in its own partition, it steals vertices
from other partitions and processes these stolen vertices.

However, a naive parallel implementation of the algorithm may have highly
skewed workloads among threads due to the power-law distribution of vertex
degree in many real-world graphs. Our algorithm only needs to perform
intensive computation ($local\_stat$ in Figure \ref{lstat}) on few vertices,
which dominates the entire computation in $top\_lstat$. Furthermore,
the time of computing $local\_stat$ on different vertices varies
significantly. Therefore, the naive load balancing scheme, which moves
the computation of an entire vertex to another thread,
is insufficient to evenly distribute the most intensive computation among threads.

Therefore, we further split computation of $\Psi_{1}(v)$ for better
load balancing by splitting $N_1[v]$ into $j$ parts
$N_{1,1}[v], N_{1,2}[v], ..., N_{1,j}[v]$. Each part $N_{1,i}[v]$ is only
responsible for computing the contribution to $\Psi_{1}(v)$ from its own part,
i.e., computing the cardinality of the intersection of $N_1[v]$ and $N_1[u]$,
for all $u \in N_{1,i}[v]$.
When load balancing is triggered, the computation of $N_{1,i}[v]$ can be moved to
another thread. Since there are many splits, each of which contains a small amount of computation, it is
much easier to distribute computation evenly among threads.

An additional issue in the parallel implementation is to maintain the maximal
locality statistic discovered currently in $top\_lstat$ without much locking
overhead. Given the fact that the maximal locality statistic is updated very
infrequently and the value increases monotonically, we always compare
a new locality statistic with the current maximal value without locking before
updating the maximal value with locking. As such, we avoid most locking for
updating the maximal locality statistic. We do not lock when we read the maximal
locality statistic. Even though we might read a stale value in a very rare case,
it does not affect
the correctness of our implementation. The worst case is that we need to
compute locality statistic on slightly more vertices.

\subsubsection{External-memory implementation}
Given a graph with billions of vertices
and hundreds of billions of edges, we can no longer store the entire graph
in RAM in a single machine. With the advance of solid state drives (SSD)
in hardware \cite{fusion} and software \cite{safs}, SSDs can now perform
over one million I/Os per second. This makes SSDs a natural extension
of RAM in large-scale data analysis, as illustrated by FlashGraph \cite{FlashGraph}.
FlashGraph stores algorithmic vertex state in RAM and edge lists on SSDs.
In order to scale, FlashGraph requires the size of vertex state to be
a small constant.

We use a very compact data structure for our algorithm to store vertex state,
which only occupies
eight bytes per vertex. The eight bytes can be used to store the locality statistic
of a vertex, the upper bound of the locality statistic, or a pointer to
the neighbor list of a vertex. We keep the neighbor list of a vertex in memory only when we
perform the expensive computation $local\_stat$ on the vertex. Therefore, we
only maintain a small number of neighbor lists in RAM at a time. Furthermore,
we read the edge lists of neighbor vertices from SSDs only when they are required.
As a result, our implementation has a small memory footprint, compared with
the graph storage size, which allows us to process graphs with billions of vertices
in a single commodity machine.

\section{Experiments}
\label{sec:experiment}
This section looks into the performance of our detection methodology on both a synthetic graph and a massive real-world graph dataset. 
To test the proposed framework on the synthetic graph, the behavior of Receiver Operating Characteristic (ROC) and Adjusted Rand Index(ARI) \cite{meilua2007comparing}\cite{zhao2012consistency}\cite{fortunato2010community} are observed under three scenarios, where $k=0,1,2$, to quantitatively evaluate how similar the partitions delivered by the framework are to the true partitions. The real-world graph used in this section is the largest Hyperlink Graph \cite{hyperlinkgraph_source} available to the public. 

\subsection{Synthetic data}
\label{sec:exp-synthetic}
The performance of our detection framework is evaluated through synthetic experiments because the underlying randomness that governs a real network is usually unknown.
The artificial graphs used in the synthetic experiments are generated from Stochastic Block Model (SBM) \cite{wasserman1994}. Note that SBM, a more generic version of Planted Partition Model \cite{PhysRevE.74.035102} \cite{condon2001algorithms}, is widely used as a testbed for community detection 
algorithms today \cite{karrer2011stochastic}\cite{fortunato2010community} \cite{zhao2012consistency} and gains the reputation of a standard benchmark. In a stochastic block model containing blocks $\{1,\dots,B\}$, $V$ is partitioned into $B$ distinct blocks $[n_1], \dots [n_B]$, where $[n_i]$ denotes the vertices in block $i$. 
The connectivity probabilities among all vertices are characterized by a $B \times B$ symmetric Bernoulli rate matrix $P$, where $P_{i,j}$ denotes the block connectivity probability between blocks $i$ and $j$. Formally, if $G$ is represented by an adjacency matrix $A$, $A_{u,v}$ is a Bernoulli random variable with success probability $P_{i,j}$ if $u\in [n_i]$ and $v \in  [n_j]$. 

In order to preserve sparsity, degree heterogeneity and built-in active community structures in a SBM graph, we select the following SBM parameter settings:
\[B=4, n_1=940, n_2=n_3=n_4=20\]
and 
\begin{equation*}
  \label{eq:P}
    \mathbf{P} = \mathbf{P}_{0}  + diag(0, 0.19, 0.29, 0.39)
\end{equation*}
where $\mathbf{P}_{0}$ is a matrix that every element is $0.01$.
Given the parameters above, $G$ is a graph having $4$ blocks where the majority block $[n_1]$ involves $94\%$ actors of the network. 
Each of the blocks $i=2$ up to $B$ has self-connectivity probability $\mathbf{P}_{i,i}=0.1\times i$. The case where $\mathbf{P}_{4,4}>\mathbf{P}_{3,3}>\mathbf{P}_{2,2} \gg \mathbf{P}_{1,1}$ is of interest because we can consider $[n_2], [n_3], [n_4]$ as three built-in active communities whose inner connectivity level is anomalously high. 


\begin{figure}[htbp]
	  	\centering
		\vspace{0mm}
		\hbox{\hspace{+4em}
	 	 \includegraphics[scale=0.4]{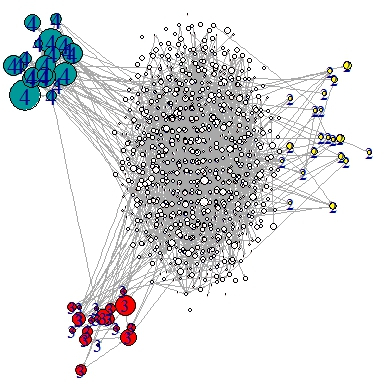}
		\vspace{-5mm}
	 }
	 	 \caption{One sample graph $G$ with $n=1000, m=10358$. One-tenth of uniformly sampled edges are incorporated in the figure. White (no label), yellow (label 2), red (label 3) and green (label 4) clusters stand for blocks $[n_1], [n_2], [n_3]$ and $[n_4]$ respectively. Size of all vertices are proportional to locality statistic $\{\Psi_{k=1}(v)\}_{v=1}^{n}$.}
	 	 \label{fig:Sampled_SBM}
\end{figure}

Figure \ref{fig:Sampled_SBM} shows a sample graph configuration of $G$ where the size of each vertex $v$ is proportional to underlying $\Psi_{k=1}(v)$.
White (no label), yellow (label 2), red (label 3) and green (label 4) clusters stand for blocks $[n_1], [n_2], [n_3]$ and $[n_4]$ respectively. We observe that sizes of vertices belonging to colored clusters are more likely to be larger than the ones in the majority white block. This phenomena foreshadows the rationale of using top $Q$ locality statistic values to cut off a massive number of negligible vertices which are unlikely to be in active communities.

The performance of separating built-in active vertices from inactive vertices by top $Q$ locality statistic values in SBM graphs is evaluated as follows.
Selection of $Q$ vertices with the largest locality statistic values to form $\mathcal{C}=\{v_1,v_2,\dots,v_Q\}$ can induce false alarms because it is likely that only a subset of $\mathcal{C}$ are 
built-in active community members in SBM random realizations. We can treat the Step (i) as a binary classification task, where $[n_2]\cup [n_3] \cup [n_4]$ are underlying positive labels and $[n_1]$ are negative ones, by using the $Q$-th largest locality statistic as a decision boundary. Next, the performance of the classifier is empirically evaluated through Receiver Operating Characteristic (ROC) curve and Area Under Curve (AUC). The empirical ROC curve is built through Monte Carlo simulations. Specifically, we repeatedly generate stochastic block model graphs and run Step (i) by varying $Q$ from $1$ to $n$ for each graph. Accordingly, we calculate true and false positive rates according to true labels for each $Q$ in each run.
\begin{figure}[htbp]
	\centering
	\vspace{-3mm}
	\hbox{\hspace{0em}
		\includegraphics[scale=0.57]{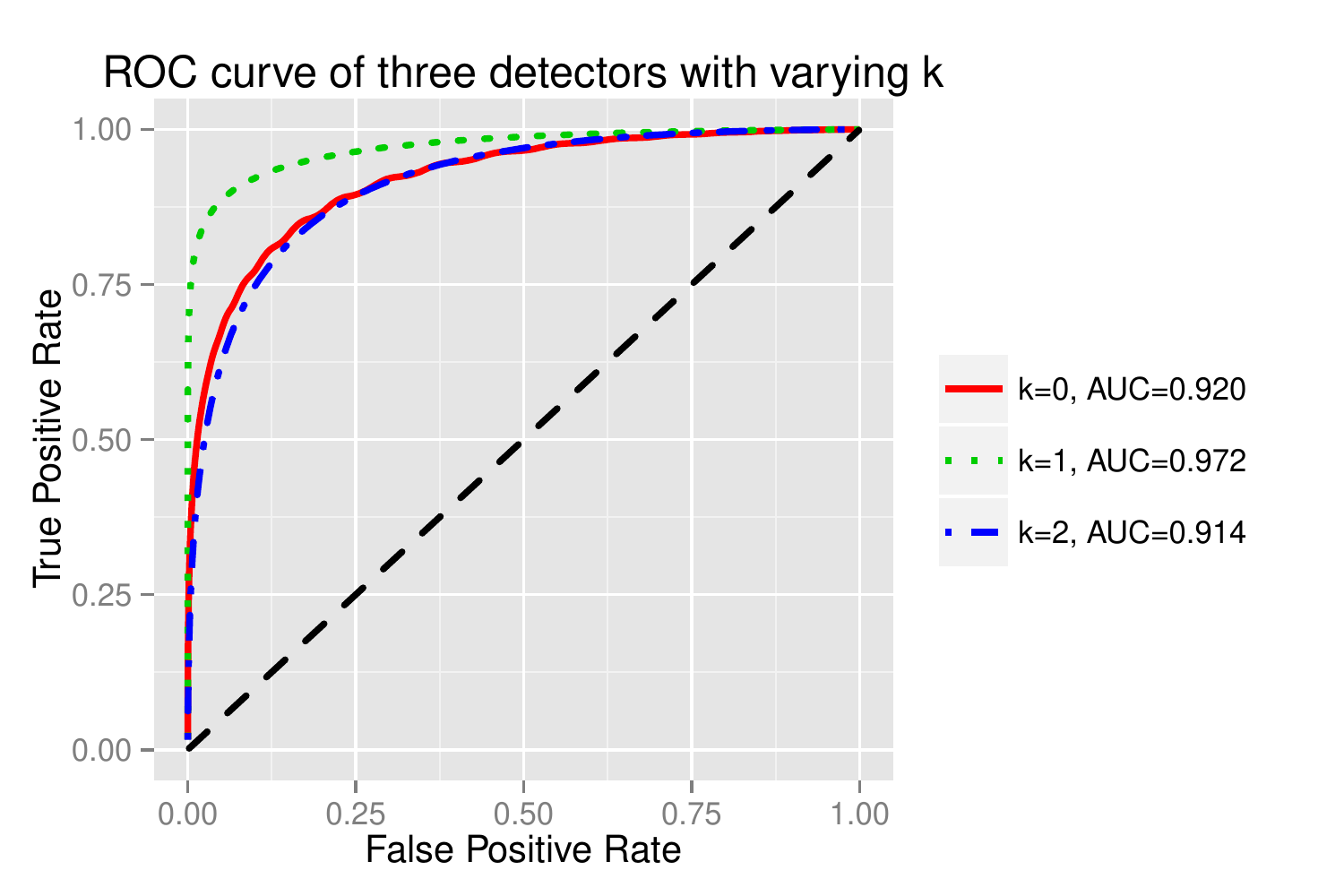}
	\vspace{-4mm}
	}
	\caption{ROC mean curves and corresponding AUCs of classifying active vertices using $Q$-th largest $\Psi_{k}(v)$ as decision boundary. The curve is built on 4000 Monte Carlos where each run generates a SBM graph and calculate one discrete ROC curve by enlarging $Q$ to increase false positive rate.}
	\label{fig:ROC}
\end{figure}

Figure \ref{fig:ROC} shows the ROC mean curve of the classifiers with different $k$ based on $4000$ Monte Carlos. All three classifiers achieve AUC over $0.9$ in this scenario, which demonstrates the usefulness of applying the $Q$-th largest locality statistic as a classifier boundary. It is also interesting to note that $\Psi_{k=1}(v)$ outperforms $\Psi_{k=0}(v), \Psi_{k=2}(v)$ in
this moderate scale graph. In a graph at this scale, compared with $\Psi_{k=0}(v)$, $\Psi_{k=1}(v)$ aggregates more edges in a larger neighborhood to outclass itself from other vertices if $v$ is in an active community. Compared with $\Psi_{k=2}(v)$, $\Psi_{k=1}(v)$ dominates in this experiment because $N_{2}(v)$ are more likely indistinguishable and highly overlapped between vertices from majority groups and active groups. 

Next, after pinning down the $Q$ most active vertices, we construct their similarity matrix $\mathcal{S}$ through the Jaccard Index in \S~\ref{jaccard_index}, and perform a classic spectral clustering algorithm with Radial Basis Function (RBF) Kernel on $\mathcal{S}$ to cluster the $Q$ vertices. This is a clustering task so that Adjusted Rand Index (ARI), recommended in \cite{meilua2007comparing}\cite{zhao2012consistency}\cite{fortunato2010community}, is an appropriate ad-hoc assessment of detection accuracy because the underlying cluster labels of the $Q$ vertices are known. 

\begin{figure}[htbp]
	  	\centering
		\vspace{-2 mm}
		\hbox{\hspace{-1em}
	 	 \includegraphics[scale=0.52]{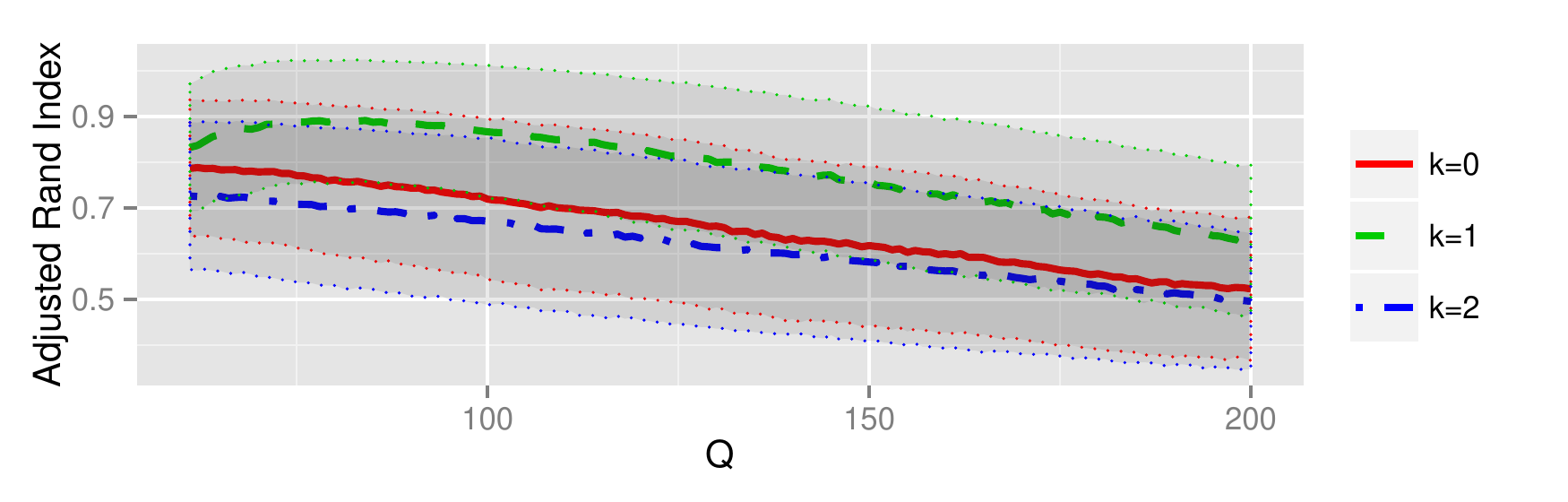}
		\vspace{-4 mm}
	 }
	 	 \caption{Adjusted Rand Index curves against $Q$, based on $4000$ Monte Carlos, between spectral clustering results and true clusterings of top $Q$ vertices.}
	 	 \label{fig:ARI}
\end{figure}

Figure \ref{fig:ARI} shows the ARI curves against $Q$, based on $4000$ Monte Carlos, between our spectral clustering results and true clusterings of the top $Q$ vertices. The horizontal axis starts from $Q=61$ to guarantee that the top $Q$ vertices precisely come from $4$ distinct clusters $[n_1], [n_2], [n_3], [n_4]$. The bold curves are mean values and dot curves are mean curves plus (or minus) one standard deviation. It is clear that all mean ARI curves are still greater than $0.5$ even when one-fifth of $V$ are classified as active community members in step (i). In fact, if $Q$ is well specified by a user, e.g., $Q<75$, ARI values of clustering based on all three locality statistics are greater than $0.7$. The results here suggest the satisfying accuracy of our detection framework.
\subsection{Real data}\label{sec:page-graph data}
In this section, we evaluate our framework on the Hyperlink graph from August
2012 Common Crawl Corpus \cite{hyperlinkgraph_source}, the largest real-world
graph dataset publicly available so far. The Hyperlink graph provides three different levels of aggregations on the graph.
In this work, we use the Page-level version of the Hyperlink graph, where each vertex is a single web page, to verify the scalability of our detection framework. The Hyperlink graph is an unweighted and directed graph with $3,563,602,789$ vertices and $128,736,914,167$ edges.
It is infeasible to perform any community detection algorithms with the complexity of $O(nm)$ or $O(n^2)$ on this graph. Furthermore, in the web graph society, a typical motivation of investigating community detection is to identify link farms which are deliberately created to increase search engine ranks \cite{fortunato2010community}. With this motivation, observers are concerned only with communities consisting of active hyperlinks. These two constraints are the obstacles of deploying other algorithms but bypassed by our detection framework.

\subsubsection{Active communities of Hyperlink Graph} We run our detection
framework on the Hyperlink graph to determine its effectiveness on the massive graph. In our experiment, we select  $k=1$ and run the trimming algorithm to identify the top $Q$ vertices of the largest locality statistic values, where $Q=2000$. In Step (ii) of the detection framework, Jaccard Index is selected to construct the similarity matrix $\mathcal{S}$ among the top $2000$ vertices. Next, to cluster pinpointed websites into active communities, we use the same spectral clustering method with RBF kernel in \S \ref{sec:exp-synthetic}. The number of clusters is suggested by the spectral gaps of $S$.
\begin{figure}[htbp]
		\vspace{-3 mm}
		\hbox{\hspace{2em}
	 	 \includegraphics[scale=0.3]{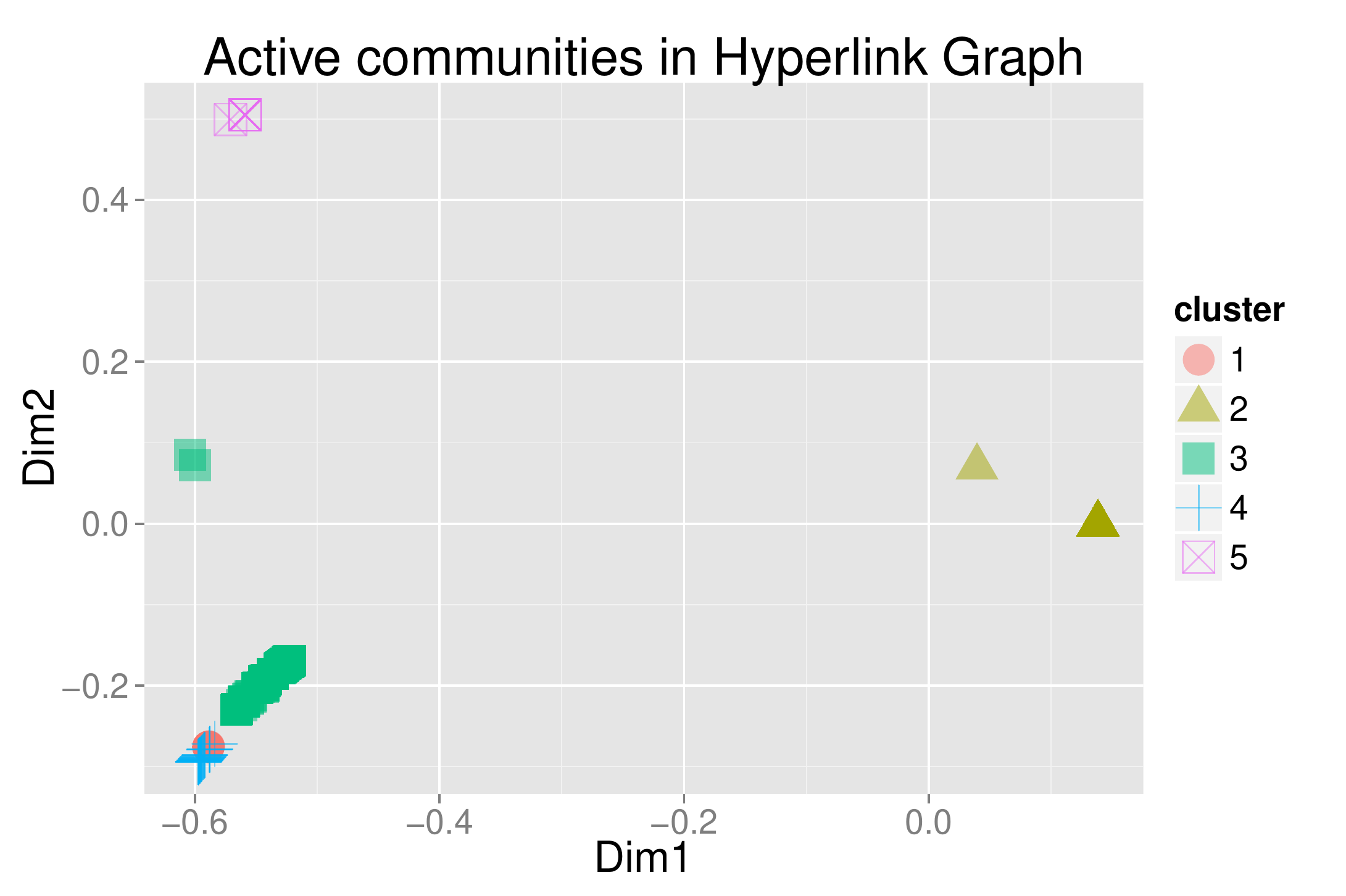}
		\vspace{-4 mm}
	 }
	 	 \caption{Five active communities in HyperLink graph. Top $Q=2000$ vertices projected into first two dimensions of classic multidimensional scaling of $\mathcal{S}$. $5$ Communities are colored separately where community index is consistent with Table \ref{table:ex-websites}. The sizes of Active community 1 to 5 are $n_1=35$, $n_2=1603$, $n_3=199$, $n_4=42$ and $n_5=121$ respectively.}
	 	 \label{fig:DensestCommunityInHyperlink}
\end{figure}

\begin{small}
\begin{table}[t]
	\footnotesize
	\begin{tabular}{ | l | l |}
    \hline
    Community  & Selected URLs \\ \hline
    1 & 	http://www.families.com/ \\
	  &  	 	 http://www.eromance.com/ \\
		&  	http://www.freecoupons.com/ \\
		&  	http://www.networkmedia.com/ \\
		&  	http://www.younger.com/ \\
\hline
    2 &  	all in this pattern:\\
			& http://www.alphateenies.com/movies/* \\ \hline
   3  & 	http://wordpress.org/ \\
		&  	http://www.google.com/ \\
		&  	http://www.flickr.com/ \\
		&  	http://www.facebook.com/ \\
		&  	http://twitter.com/ \\ \hline
     4  &  	http://www.amazon.com/ \\
		&  	http://www.zappos.com/ \\
		&  	http://www.abebooks.com/ \\
		&  	http://www.myhabit.com/ \\
		&  	http://www.woot.com/ \\
\hline
    5 &  	http://www.acidmovies.com/   \\
		&  	http://www.azimuthmovies.com/   \\
		&  	http://www.drymovies.com/   \\
		&  	http://www.btwmovies.com/   \\
		&  	http://www.effectmovies.com/ \\ \hline
    \end{tabular}
	\caption{Table of five selected URLs from active communities in Hyperlink Graph
		provided by our detection framework. URLs of similar topics are clustered
		in the same active communities. Community 1 are URLs maintained and
		developed by \textbf{networkmedia} company; Community 2
		and 5 are collections of adult websites; Community 3 are popular social
	media sites. Community 4 are online shopping sites.}
	\label{table:ex-websites}
\end{table}
\end{small}
The procedure above detects five colored active communities decomposed from $2000$ vertices (Figure \ref{fig:DensestCommunityInHyperlink} and Table \ref{table:ex-websites}).
In Figure \ref{fig:DensestCommunityInHyperlink}, the top $2000$ vertices are projected into a two-dimensional space through classical multidimensional scaling (MDS) on the similarity matrix $\mathcal{S}$. Five active communities obtained from our detection framework are colored separately. The sizes of community 1 to 5 are $n_1=35$, $n_2=1603$, $n_3=199$, $n_4=42$ and $n_5=121$ respectively. Table \ref{table:ex-websites} lists five selected web URLs from each cluster for further illustration of detected communities.
Out of $2000$ vertices, there are $1603$ vertices forming the community 2
whose members are all hyperlinks extracted from a single Pay-level-domain adult website
(i.e. \url{ http://www.alphateenies.com}). Community 1 is a collection of websites
that are all developed, sold or to be sold by an Internet media company \textbf{networkmedia}.
Community 4 consists of websites related to online shopping such as the shopping giant Amazon and the bookseller AbeBooks. Community 5 is another collection of 121 adult web pages where each web page comes from a different Pay-level-domain in this cluster. In the community 3, most links are social media websites and often used in our daily life such as WordPress.org and Google. In summary, top 5 active communities in the Hyperlink Graph are grouped with high topical similarities, which is consistent with findings in \cite{fortunato2010community}. Therefore, these noteworthy clusters produced by our detection framework not only imply its applicability on a massive graph but also practicality on real World Wide Web graphs.

\subsubsection{Time-saving trimming algorithm}
\begin{figure}[htbp]
	  	\centering
		\vspace{-0mm}
		\hbox{\hspace{3em}
	 	 \includegraphics[scale=0.41]{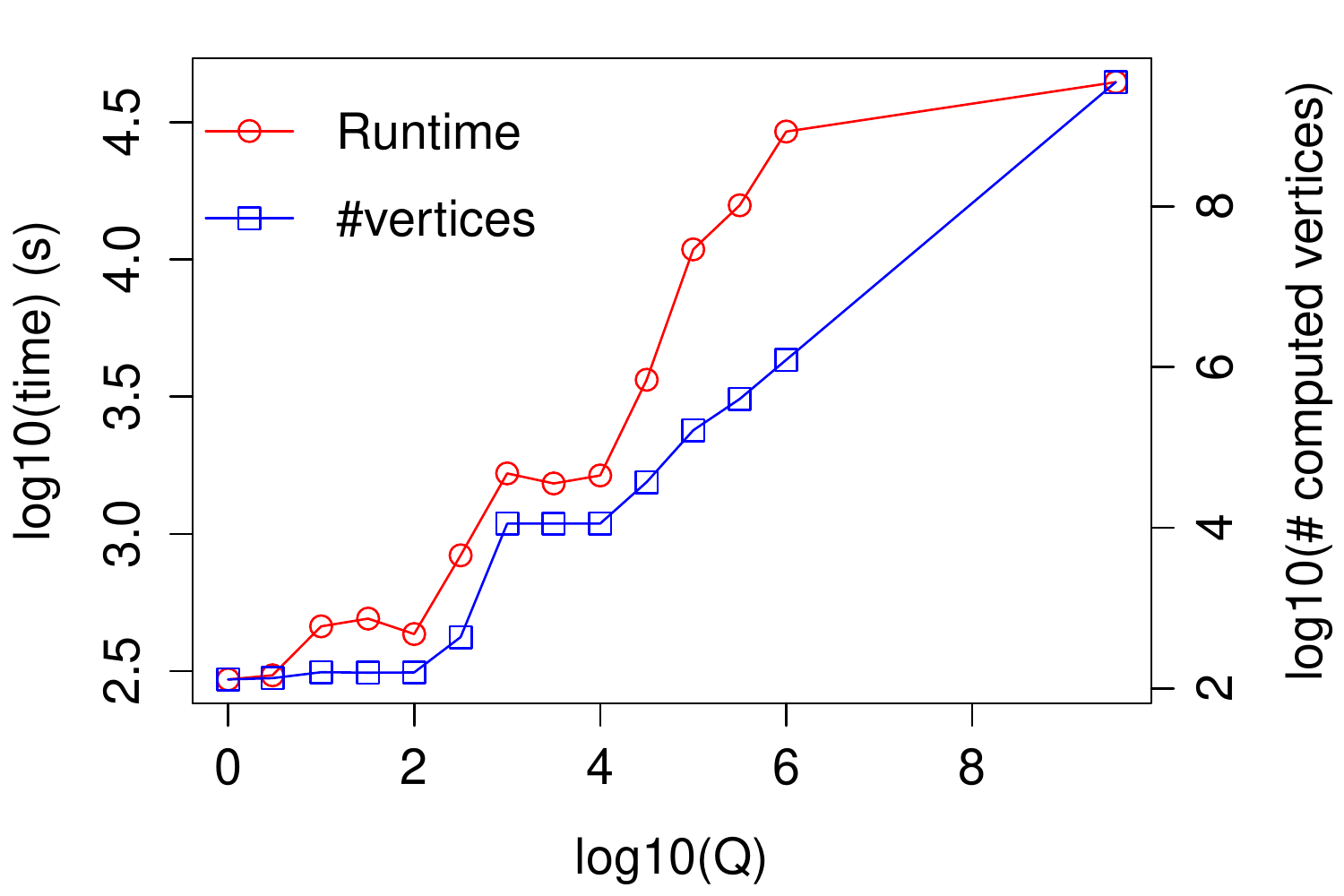}
		\vspace{-4mm}
	 }
	 	 \caption{Log-log plot of time consumption and the number of locality statistic-computed vertices against $Q$ of trimming algorithm. The log base is $10$ and $Q$ ranges from $1$ to $n$. In the Hyperlink graph, the running time of trimming algorithm $T(Q) = O(\sqrt{Q})$ and computing top $Q=10^4$ locality statistic values only takes $3.7\%$ time consumption on all locality statistic values}
	 	 \label{fig:log-log}
\end{figure}
We evaluate the time saving achieved by the trimming algorithm (Section \S \ref{sec:local.scan-implement}) on the massive Hyperlink graph.
The computing environment of conducting the trimming experiment is a machine with four Intel Xeon E5-4620 processors, clocked at 2.2 GHz, and 512 GB memory of DDR3-1333. Each processor has eight cores with hyperthreading enabled, resulting in 16 logical cores. The machine has three LSI SAS 9207-8e host bus adapters (HBA) connected to a SuperMicro storage chassis, in which 12 OCZ Vertex 4 SSDs are installed.
We conduct an experiment to show the relation of time consumption against $Q$ and the number of locality statistic values computed against $Q$ in trimming.

This experiment demonstrates that the trimming algorithm winnows active vertices efficiently even if $Q$ is large. Figure \ref{fig:log-log} shows that the running time of computing top $Q$ locality statistic values on the Hyperlink graph is sub-linear against $Q$.
For example, the ratio $T(Q=10^4)/T(Q=n)=0.03689694$ implies that computation on top $Q=10^4$ vertices only takes $3.7\%$ time of computation on all vertices because our algorithm only needs to compute locality statistic on $0.00032\%$ vertices in the graph to find top $10^4$ vertices.

\section{Conclusion}
\label{sec:discussion}
In this paper, we propose a novel framework for detecting active commmunities that scales to a billion-node graph. Our framework consists of two parts: trimming of inactive vertices and clustering on selected active vertices. In the trimming step, we employ the locality statistic and present a parallelizable algorithm to distribute computation. 
The results on synthetic SBM graphs indicate that our framework performs well and yield reasonable active communities. A general strength of our method is that, unlike most other approaches, it is scalable to extremely massive graphs. Its application to the Hyperlink graph with billions of vertices discovers meaningful communities in the real World Wide Web graph dataset. 

There are some future research avenues open to this work. 
For future work, if $G$ is weighted, $\Psi_{k}(v)$ can be extended to the sum of edge weights in the subgraph of $G$ induced by $N_{k}[v]$. Another research direction is to study the optimal choice of $k$ as varying $k$ may yield different trimming results. We should also explore the trade-off between heavier trimming computational burden and trimming performance. Finally, although our current experiment uses a combination of Jaccard Index and spectral clustering to perform clustering, it might be interesting to figure out if there is an alternative combination dominating our current approach. 
																																							


\bibliographystyle{siam}
\bibliography{WZBP}

\end{document}